# On the General Nature of Physical Objects and their Interactions, as Suggested by the Properties of Argumentally-Coupled Oscillating Systems

Danil Doubochinski and Jonathan Tennenbaum

*Quantix - Société de Recherche et Développement en Technique Vibratoire*
*86, Rue de Wattignies, 75012 Paris, France*
doubochinski@hotmail.com ; tennenbaum@debitel.net

The work reported here originates in the discovery, four decades ago, of a previously-unknown type of self-organizing interaction among oscillating systems -- so-called argumental interactions -- and of "quantized" modes of behavior in macroscopic argumentally-coupled oscillators, having no equivalent in the classical theory of oscillations [4-18]. Argumental interaction is characterized by the property, that the exchange of energy is regulated by phase-frequency-amplitude fluctuations in the participating oscillating systems, while each of them operates at very nearly its own proper frequency and retains (in the mean) its characteristic dynamic parameters. Argumental interactions can be demonstrated in a variety of electromechanical devices, the simplest of which belongs in every physics classroom: A pendulum (1 Hz) interacting with an alternating-current electromagnet (30-1000 Hz), and possessing a discrete series of stable amplitudes. The argumental mechanism lies at the basis of a second remarkable phenomenon, discovered around the same time: when placed in a high-frequency EM field, electrical resonators (such as LCR circuits), coupled to each other by inductive, capacitative and resistive couplings and free to move in space under the influence of the corresponding ponderomotive forces, show the tendency to group themselves into stable spatial configurations [19-24].

Argumental interactions were the subject of extensive experimental, numerical and theoretical investigations during the 1970s and 1980s, reported mainly in Soviet physics journals. Among other things, research revealed striking similarities between the behavior of argumentally-coupled macroscopic oscillators, and the quantum behavior of atoms and other microphysical objects. At the same time it was demonstrated that argumental interactions can provide an efficient mechanism for the self-organizing coupling and transfer of energy between oscillating systems whose frequencies can differ by several orders of magnitude. These results, while curiously little-known in the scientific community today, have led to some significant technological developments of current interest [25-29].

In recent years the present authors have been jointly pursuing new lines of investigation into argumental interactions and their possible significance for the foundations of physics. Among other things, the study of argumentally-coupled oscillators suggested to the authors a new general way of looking at physical objects, their interactions and their aggregative ("social") behavior as manifested on all scales of observation. We believe this new viewpoint, which differs significantly both from that of classical physics and from quantum theory as presently understood, might lead to a more coherent understanding of many natural phenomena which until now have been studied only in a piecemeal fashion and from divergent points of view in physics, chemistry, astronomy and biology. We consider it a strength of our approach that it arose not from abstract speculation, but from the discovery of concrete physical phenomena.

In order to make the authors' train of thought intelligible to the reader not familiar with the field of argumental interactions, we have combined the presentation of our ideas with an extended review of experimental and theoretical results on argumentally-coupled oscillators and the associated "Macroscopic Quantization Effect" (MQE). Our review serves the additional purpose of introducing this little-known, but most interesting field of research to a broader scientific audience.

Our paper is organized as follows. The first two sections deal with the phenomenology of argumentally-coupled systems and the physical mechanism underlying the emergence of "quantized" oscillatory regimes in such systems, using the argumental pendulum as an illustration. The discussion of the pendulum may seem a bit tedious to some readers, but is indispensable for gaining a physical grasp of how argumental coupling actually works. The third section summarizes results concerning another specific case of great theoretical interest: the argumental analog of the "elementary oscillator" which Max Planck employed in his original investigations on blackbody radiation. These results suggest a natural physical mechanism for the emergence of discrete energy states in the interaction between an oscillator and an electromagnetic wave, independent of the assumptions of quantum mechanics. In the fourth section we discuss a more general mechanism for the emergence of "quantized" regimes, which arises from the interplay of electromagnetic and mechanical oscillations, and manifests itself in the aggregative ("social") behavior of multiply-coupled oscillating systems. In the fifth section we elaborate on the "participatory-generative" character of argumental coupling, contrasting this to the forms of coupling commonly studied in classical as well as quantum physics. We argue that each of the quasi-stationary regimes of argumentally-coupled oscillators deserves to be regarded as a real physical object in its own right, and show how argumental interactions provide a mechanism for generating an entire hierarchy of distinct physical objects, starting from a given set of oscillators. The final section presents our ideas concerning the foundations of physics, suggested by the results presented in the preceding sections.

**1. The phenomenology of argumentally-coupled oscillators**

Argumental interactions and the associated "Macroscopic Quantization Effect" were originally discovered in 1968 by Danil and Yakov Doubochinski, then students at Moscow University. The classical example is the so-called argumental pendulum (Fig. 1): A low-friction (high Q-value) pendulum with a natural frequency of 0,5 - 1Hz and a small permanent magnet affixed to its end, interacts with the magnetic field produced by a narrow solenoid (with air core) located under the pendulum's equilibrium position and fed by higher-frequency alternating current of between 30 and 1000 Hz. The coupling of the pendulum with the magnetic field results in an oscillating system having a discrete array of stable regimes, each corresponding to a specific amplitude of quasi-stationary oscillation of the pendulum. In each of those regimes the pendulum oscillates with a frequency near to its undisturbed proper frequency.

An essential condition for this "quantization" phenomenon is the strong *spatial inhomogeneity* of the magnetic field of the solenoid: the field exerts a significant influence on the pendulum only within a narrow "interaction zone" around the solenoid, outside of which the field strength drops rapidly to zero. This inhomogeneity permits the pendulum to regulate its own exchange of energy with the alternating field via small shifts (fluctuations) in its phase relative to that of the field.

The differential equation describing this system has the property, that the angular coordinate of the pendulum appears in the argument of the function defining the force acting on the pendulum. This was the original reason for the terms "argumental interaction" and "argumental coupling", which came to be applied to a whole class of oscillating systems, sharing a common mechanism of self-regulation via phase-frequency-amplitude fluctuations.

Setting the pendulum into motion, the following behavior is easy to observe.

1. When released from any given position, the pendulum's motion evolves into a stable, very nearly periodic oscillation, whose amplitude belongs to a *discrete set of possible values* (Fig. 2).

2. In each of the stable modes, the pendulum's frictional losses over a given period of oscillation are compensated in a self-regulated fashion by a transfer of energy from the field of the solenoid, thereby sustaining the pendulum's oscillation in a quasi-stationary regime near to its own natural frequency.

3. The number of available stable oscillation regimes and the values of the corresponding amplitudes, depend strongly on the frequency of the alternating current in the solenoid. The higher the frequency, the larger the number of stable regimes that can be excited. The stable regimes correspond to periods of the pendulum which are close to an odd-number integral multiple of the period of the alternating magnetic field, and at the same time not far from the pendulum's proper, undisturbed period.

4. The pendulum's motion in a given regime is never strictly periodic, but constantly fluctuates slightly in phase, frequency and amplitude around certain stable average values. These fluctuations are essential to the mechanism of stability of the quantized oscillatory regimes.

5. The values of the "quantized" stable amplitudes are practically independent of the strength of the alternating current supplied to the electromagnet, over a very large range. The pendulum compensates for changes in the strength of the magnetic field, by slightly adjusting its moments of entry into the zone of interaction (i.e. its phase relative to that of the alternating current supplied to the solenoid), while maintaining almost exactly the same average amplitude and frequency. If we gradually reduce the strength of the current in the electromagnet, we reach a threshold below which the given regime can no longer support itself, and the pendulum decays into a lower-energy state.

In an impressive variant of the pendulum experiment, a whole array of pendula of different lengths is maintained in quasi-stationary oscillation, each near its own natural frequency and at any one of its own discrete set of stable amplitudes, by interaction with the field of a single elongated solenoid fed with alternating current of a single frequency (Fig. 3).

Several other types of argumental oscillators with quantized amplitudes can be obtained as variations on the argumental pendulum. For example, if we replace the pendulum by a rotating horizontal bar, we obtain an "argumental rotor" having a discrete set of quasi-stationary rotation speeds. Another interesting variant is to replace the pendulum by a quasi-linear mechanical oscillator, of the type of a balance wheel of a clock.

A particularly important example, difficult to realize experimentally but of great theoretical interest, is the argumental analog of the "elementary oscillator" which Max Planck employed in his original investigation of the spectrum of blackbody radiation [1]: A mass carrying an electric charge, fixed at the end of a spring and free to oscillate in the x-direction, interacts with an electromagnetic wave-field propagating along the same axis (Fig. 4). Investigations by mathematical and computational methods showed that when the wave frequency is large compared to the proper frequency of the oscillator, the interaction of the oscillating charge with the longitudinal force exerted by the wave field, automatically gives rise to a discrete array of stable amplitudes of the oscillator – independently of any quantum mechanical assumptions [14].

Theoretical and numerical studies also showed, that the effect of quantization of amplitudes in the argumental interaction between a mechanical oscillator and an electromagnetic wave does not depend upon the presence of dissipation or explicit nonlinearity of the oscillator, but would occur even for an ideal frictionless linear oscillator. This is an important result. The quantized regimes arising in argumental interactions do not represent "dissipative structures" in the sense of Prigogine, and should also not be confused with forms of self-organization that depend upon strong nonlinearity of the component systems. What is essential to the genesis of discrete regimes in argumentally-coupled systems, is the establishment of a *specific form of "dialog" between the interacting systems*: a dialog realized through constantly fluctuating, phase-frequency-amplitude-modulated exchange and transformation of energy. This dialog provides the self-regulating mechanism maintaining each of the quasi-stationary regimes of the coupled system.

The argumental mechanism just described, gives rise to three unique properties of argumental interactions, to which the authors attach *fundamental* significance. These properties are essential to the ideas we shall develop in the final section of this paper:

*In each of the stable regimes of an argumentally-coupled system (1) the participating oscillators retain (in the mean) very nearly their original frequencies and other oscillatory characteristics as in the uncoupled state, while at the same time (2) the coupled system constitutes an individual dynamical object with distinct characteristics and parameters of its own, generally very different from those of the component systems, and (3) a single oscillator may simultaneously participate in many coupled systems.*

## 2. The physical mechanism of argumental coupling and the Macroscopic Quantization Effect (MQE)

The basic principles of argumental interactions emerge most clearly when we examine the case of the argumental pendulum using two different, but complementary idealizations. The first exhibits most simply the phase-dependent nature of argumental interactions and the genesis of "quantized" amplitudes; the second focusses on the mechanism of resonant transfer of energy between widely differing frequencies and points the way to a general theory of argumental interactions.

The first approach begins by assuming that the magnetic field of the solenoid is practically homogeneous within a certain narrow "interaction zone" around the vertical, equilibrium position of the pendulum near the solenoid, and drops off "instantly" to zero outside that zone (Fig. 5). The pendulum's interaction with the alternating magnetic field can then be described in the following straightforward manner.

For small initial amplitudes, for which the pendulum remains entirely within the interaction zone around the electromagnet, the external force acting on the pendulum is practically independent of its position and has the form

$F = F_0 \cos(\nu t)$

where $\nu$ is the frequency of the alternating current supplied to the solenoid. Here we have the classical case of damped forced oscillations under an external periodic force. Given the large difference between $\nu$ and the proper frequency of the pendulum, the classical analysis shows that long-period motions of the pendulum will be damped out, leaving only small vibrations around the equilibrium point at the frequency of the external field.

An entirely different situation arises when the initial motion of the pendulum is large enough to carry it beyond the limits of the "interaction zone". In this case the external force on the pendulum can be represented by the function

$F = F(X, t) = e(X) F_0 \cos(\nu t)$

where X is the angular position of the pendulum and $e(X)$ is the function defined by $e(X) = 1$ when $-x_0 < X < x_0$ (interaction zone) and $e(X) = 0$ outside that interval. Now examine the pendulum's motion in terms of the successive half-periods between the maximum heights of the pendulum (i.e. the moments at which its velocity is zero). In each such half-period, the pendulum descends in free motion from its maximum to the point where it enters the interaction zone. Transiting through the zone of interaction, the pendulum experiences alternately accelerating and decelerating impulses from its interaction with the alternating magnetic field. The net gain or loss of energy of the pendulum as a result of its transit through the interaction zone, depends upon the *phases* of the solenoid current at the moment of entry and at the moment of exit. If the *transit time* corresponds to an *integral number of periods of the electromagnet*, then the effects of the accelerating and decelerating half-periods of the field will cancel out; but if the pendulum leaves the zone of interaction after a *non-integral* number of periods of the electromagnet, relative to the moment of entry into the zone, then the pendulum will experience a net non-zero accelerating or decelerating effect from its interaction with the alternating field. Exiting from the interaction zone, the pendulum ascends freely to a new maximum height (in general different from the preceding one), and thereafter descends again in its next half-period.

It is easy to see that the functional relationships between the phases and amplitudes for successive half-periods are rather complicated, and the attempt to characterize the system's long-term evolution in the general case poses enormous difficulties. But a direct analysis is possible in the special case of *quasi-stationary* regimes, whose existence is known from experiments on real argumental pendula.

Assuming a motion of very nearly fixed amplitude, we can regard the exchange of energy between the pendulum and solenoid as determined essentially by *phase relations* alone. The existence of stable regimes depends upon the form of the functional relationship between the phase of entry of the pendulum into the interaction zone (for a given amplitude) and the net acceleration (or deceleration) experienced by the pendulum in traversing the zone. Although it is possible to establish the *existence* of stable quasi-stationary regimes of the argumental pendulum by direct analysis of this functional relationship, the method is tedious and has the major disadvantage, that it does not easily extend to the many other systems in which the Macroscopic Quantization Effect can be demonstrated. This disadvantage is remedied by the second approach, which we shall present in a moment.

It is, however, quite easy to state a *necessary* condition for a *stationary* regime, which implies that the corresponding amplitudes must belong to a *discrete set of values*: In each half-period of a stationary regime, the pendulum must receive the same portion of energy from the field of the solenoid as in the preceding half-period. Bearing in mind that the pendulum's direction of motion reverses every half-period, this means the phase of the alternating current in the solenoid must be exactly the opposite from the preceding moment of entry. Thus, the time between successive entries into the interaction zone must have the form $(m + \frac{1}{2})\tau$, where $\tau = 1/\nu$ is the period of the solenoid current, and m is some whole number. Twice that time is equal to a full period T of the pendulum, so we get a relationship $T = (2m + 1)\tau$. Thus, the possible periods of stationary regimes of the argumental pendulum are contained in this discrete series of values. Since the period T of a circular pendulum increases as a function of its amplitude (anisochronicity), and since $\tau$ is small compared to the pendulum's proper period, there will in general exist a series of amplitudes for which T coincides with an odd-number multiple of $\tau$. These amplitudes can be calculated using the classical formulas for the dependency of the period from the amplitude in a circular pendulum, assuming the pendulum's anisochronicity is not greatly altered by its interaction with the magnetic field. The values of the discrete amplitudes, estimated in this way, are found to be in approximate agreement with those of the stable regimes, actually observed in experiments. In practice the motion is never strictly periodic, but "wanders" around the values of phase, period and amplitude corresponding to a stationary regime.

This analysis makes quantization appear to depend on the nonlinearity and anisochronicity of the pendulum. It is an experimental fact, however that the phenomenon of quantized stable amplitudes persists – albeit with different numerical values for the stable amplitudes -- even when the pendulum is replaced by a quasi-linear mechanical oscillator (such as the balance-wheel of a clock) whose anisochronicity is extremely small. In this case quantization is connected with variations in the transit time through the zone of interaction as a result of the oscillator's interaction with the magnetic field.

We now turn to a second approach to the analysis of the argumental pendulum, which is much more broadly applicable and in fact leads to a general theory of argumental interactions. The second approach begins with the idea of describing the pendulum's interaction with the field of the solenoid in terms of an amplitude-frequency-phase modulation of its original free motion.

For this purpose it is useful to abstract from the specific features of the pendulum, and consider an arbitrary one-dimensional mechanical oscillator, linear or nonlinear, that possesses a family of periodic motions, depending on initial conditions, each associated with a specific amplitude, frequency and relative phase. Choosing a suitable generalized coordinate X for the oscillator, we represent the family of motions by a function $X = X(A, \omega, \phi, t)$, such that for each set of values $(A, \omega, \phi)$, X constitutes a periodic function in t with frequency $\omega$, amplitude A, and phase factor $\phi$. We think of this function as representing the free motions of the mechanical oscillator. Now suppose the our mechanical oscillator interacts with a periodic, spatially-inhomogenous field whose frequency $\nu$ can be orders of magnitude higher than the frequency range of the mechanical oscillator, and that the effect of the field on the motion of the mechanical oscillator can be described by a differential equation of the form

$$X'' + 2\beta X' + g(X) = f(X) \cos(\nu t) \qquad (1)$$

This equation includes the argumental pendulum as a special case, namely when $g(X) = \omega_0^2 \sin X$, f(X) is the step-function $e(X)F_0$ defined above, and $\beta$ is the coefficient of dissipation of the pendulum. As it turns out, the Macroscopic Quantization Effect itself does not depend strongly on the particular form of the function f that expresses the spatial dependence of the field's action. The form of the function does determine the numerical values of the "quantized" amplitudes, in a somewhat analogous fashion to the way the form of the "potential wall" determines the values of discrete energy levels of a bound particle in quantum mechanics. (The right side of the equation (1) can be replaced by a more general function of X and t, but this is not of importance for the present discussion.)

We next assume the essential characteristics of the mechanical oscillator are not greatly changed by the interaction with the field, so that its motion can be closely approximated by a "free motion" $X = X(A, \omega, \phi, t)$, in which the parameters $A, \omega, \phi$ are no longer strictly constant, but can fluctuate slowly in time around certain average values.

Turning to the right side of equation (1), we can now see how the mechanical oscillation "modulates" the external field to generate new spectral components of the field's action. For any fixed values of $A, \omega, \phi$, the function f(X) is a periodic function of t; it can therefore be represented by a Fourier series in t, whose coefficients are functions of $A, \omega, \phi$:

$$f(X(A, \omega, \phi, t)) = \Sigma c_n(A, \omega, \phi) e^{in\omega t} \qquad (2)$$

We can now express the right side of equation (1) as:

$$[\Sigma c_n(A, \omega, \phi) e^{in\omega t}] \cos(\nu t) = \tfrac{1}{2} \Sigma c_n(A, \omega, \phi) e^{in\omega} [e^{i\nu t} + e^{-i\nu t}]$$
$$= \tfrac{1}{2} \Sigma c_n(A, \omega, \phi) e^{i(n\omega + \nu)t} + \tfrac{1}{2} \Sigma c_n(A, \omega, \phi) e^{i(n\omega - \nu)t} \qquad (3)$$

From this we see that the "force term" on the right side of the equation (1) contains components of frequencies $n\omega + \nu$ and $n\omega - \nu$, where n can be any integer, positive, negative or zero. Regardless of how large $\nu$ may be in comparison with $\omega$, this spectrum will contain components of the same order as ω. Particularly interesting is the case in which ν is equal to an integral multiple of ω. Setting ν = Nω, it is easy to see that (3) will contain two terms whose frequencies are *exactly* equal to ω, namely for n = 1-N in the first sum and n = N+1 in the second sum. Taken together, these terms produce a component

$$[\, c_{1-N}(A, \omega, \phi) + c_{N+1}(A, \omega, \phi)\,]\, e^{i\omega t} \qquad (4)$$

of frequency ω. The presence in the spectrum of the force experienced by the oscillator, of a spectral component whose frequency coincides with that of the oscillator itself, opens up the possibility of a resonant transfer of energy from the high-frequency field to the low-frequency oscillator. This is exactly what occurs in the stable regimes of the argumental pendulum and related systems.

It is most important to note that the low-frequency component (4) did not originally exist in the field, but was generated by its interaction with the mechanical oscillator, in a manner akin to frequency-phase modulation in radio communications. The self-organization of *resonant transfer of energy* via the *generation and selection of additional frequency components*, is a characteristic feature of argumental interactions. One could say that the interacting systems, through this mutual modulation, create a specific channel and find a "common language" for their communication.

To go beyond these qualitative remarks it is necessary to compare both sides of equation (1), focussing on fluctuations in the values of A, ω and φ, which provide the regulatory element permitting the establishment and maintenance of stable, quasi-stationary oscillatory regimes. Analysis leads to a complicated set of equations involving the coefficient functions. With the help of the averaging method of Krylov-Bogoliubov-Mitropolski it is possible to show the existence of quasi-stationary regimes corresponding to a discrete series of average values of (A, ω, φ). For details we refer the reader to the literature on the mathematical theory of argumental oscillations, developed in the Soviet Union during the 1970s and 1980s and presented in a series of papers and conference presentations [7,12,14,16]. This theory was able to account for most of the qualitative features of argumental oscillations, and showed good quantitative agreement with experiments and numerical simulations. It also led to practical algorithms for calculating the values of the quantized amplitudes for the pendulum and other systems operating on the argumental principle.

In addition, these investigations established a close relationship between argumental coupling and the solution of certain types of differential equations with deviating arguments [8,9].

## 3. The argumental Planck oscillator

A particularly interesting special case dealt with by the theory of argumental interactions, is the above-mentioned argumental analog of Planck's "elementary oscillator" (Fig. 4). This system is described by the differential equation:

$$X'' + 2\beta X' + \omega_0^2 X = F_0 \sin(\nu t - kX) \qquad (5)$$

When X = A cos (ωt + ϕ), the right side of (5) has a well-known Fourier series expansion in terms of the Bessel functions:

$$F_0 \sin(\nu t - kA \cos[\omega t + \phi]) = F_0 \sum J_n(kA) \sin([\nu - n\omega]t - n\phi) =$$

$$F_0 \sum [\cos(n\phi) J_n(kA) \sin(\nu - n\omega)t - \sin(n\phi) J_n(kA) \cos(\nu - n\omega)t] \quad (6)$$

where $J_n$ is the Bessel function of order n. The analysis of the system proceeds by comparing the right and left sides of (5), assuming slowly varying values of A and ϕ, and applying the averaging method of Krylov-Bogolyubov-Mitropolski. We refer the reader to [14] for details, and state only the main conclusion here:

*For every integer N such that ν/N is sufficiently close to the proper frequency $\omega_0$ of the linear oscillator, there exist a series of stable regimes of the coupled system of oscillator and wave, for which the oscillator's frequency is $\omega = \nu/N$ and its amplitude $A_i$ is equal to $j_{N,i}$, the i$^{th}$ extremum of the Bessel function $J_N$.*

Numerical calculations have confirmed the theoretical analysis to good approximation, while at the same time demonstrated the remarkable fact, that the phenomenon of quantized amplitudes persists even when the coefficient of dissipation β of the linear oscillator is equal to zero.

We consider these results to be of fundamental importance. They suggest a natural physical mechanism for the emergence of discrete energy states in the interaction between an oscillator and an electromagnetic wave, which is independent of the postulates of quantum mechanics. *Could it be that interactions of the argumental type lie at the origin of the peculiar properties of microphysical objects, which we associate with Planck's quantum of action?* We shall not attempt to answer this question now, but permit ourselves the following remark.

Despite its enormous successes, quantum mechanics does not provide an intelligible explanation for the existence of the quantum of action itself; nor does it propose any physical mechanism behind such events as "the collapse of the wave function". Einstein, de Broglie and others hoped that the paradoxical nature of quantization and the schism between classical and quantum physics might finally be overcome, by a theory taking into account the *essential nonlinearity* of elementary physical processes. The discovery of the Macroscopic Quantization Effect in argumentally-interacting oscillators suggests that the mechanism of quantization in the microworld should be sought more in the *processes of interaction* of physical objects, than in the objects per se. Such an approach would be closer to the original standpoint of Max Planck, than the later (1905) suggestion by Einstein, according to which electromagnetic radiation should be regarded as already quantized, independently of its interaction with material systems [2].

However, at the time Max Planck carried out his fundamental investigations on the spectrum of blackbody radiation, his basic model for the interaction of matter with the electromagnetic field was an array of Hertzian oscillators exchanging energy with the field according to Maxwellian electrodynamics. The possible role of *spatial motion* of the oscillators was not explicitly considered. Indeed, the MQE and related phenomena arising from the interplay between mechanical and electromagnetic oscillations of differing frequencies, were not known at the time of Planck and Einstein. Perhaps the time is now ripe to reconsider these matters.

**4. Multiply-coupled oscillators and their grouping in stable configurations**

The argumental pendulum and the argumental analog of Planck's elementary oscillator are only special cases of a much more general type of coupling of mechanical and electromagnetic oscillations. In the same period as the initial investigations of the Macroscopic Quantum Effect, the Doubochinskis and their collaborators discovered a second fundamental phenomenon connected with this more general form of interaction: *the tendency of electrical oscillators, coupled by more than one form of electrodynamic coupling and allowed to move freely in space under the influence of ponderomotive forces between them, to spontaneously assemble themselves into stable configurations* [19-24]. The significance of this discovery can be seen from the following remarks:

The intimate interrelationship of ponderomotive and electromotive forces has been recognized for nearly two centuries and lies at the heart of electrodynamics. But as far as the authors have been able to ascertain, the implications of that ponderomotive/electromotive dualism for the interaction and coupling between *oscillatory* systems, have never been explored in a systematic and comprehensive manner.

It is well known that inductive, capacitative and resistive couplings between the components of electrical oscillating circuits give rise to mechanical forces between them, and hence also to various sorts of mechanical vibrations. In this way electrical and mechanical oscillations never exist in isolation, but are always present simultaneously. It had not been realized, however, that the interplay of these two types of oscillations leads to self-organizing behavior in oscillating systems.

Imagine for example, that two LCR-circuits, mounted on rigid platforms, are inductively coupled by placing the inductive elements of the two circuits (in the form of coils) parallel and near to each other (Fig. 6). In that case the coils will experience a varying mechanical force proportional to the product $J_1 \times J_2$ of the currents in the two loops Assuming both currents are sinusoidal functions of t with a common frequency f, it is easy to see that the net mechanical force, integrated over a single period, will be proportional to the cosine of the angular phase difference between the two currents.

Now imagine that the platforms, upon which the LCR circuits are mounted, can move freely in space. In that case the ponderomotive force between the inductive elements gives rise to mechanical motion, causing the circuits to change their relative positions. This in turn changes the value of the coefficient of mutual induction, thereby "modulating" the electrical oscillations in the coupled system. Finally, changes in the electrical oscillations lead to changes in the ponderomotive force acting between the circuits.

What we have said about inductive coupling, holds also true for resistive and capacitative forms of coupling. The feedback between mechanical motion and electrical oscillations, via variations in the coefficients of coupling and of the relative *phases* of oscillations in the interacting circuits, opens up the possibility of emergence of new forms of combined electro-mechanical oscillations and self-regulating, self-organizing behavior, which have no equivalent in classical treatments of coupled oscillating systems.

Theoretical and experimental investigations carried out by the Doubochinskis and their collaborators at the Vladimir State Pedagogical Institute [19-22], demonstrated that the simultaneous presence of *more than one dynamic form of coupling* radically transforms the behavior of the coupled system, and leads under certain conditions to a pronounced tendency for coupled oscillating systems to group together in stable configurations.

Fig. 7 depicts a typical experimental setup in schematic form. Systems $S_1$ and $S_2$ are LCR circuits, where $S_1$ is provided with a sinusoidal voltage source E of frequency f, and $S_2$ operates as a passive resonator. The two circuits are coupled to each other by inductive, capacitative and resistive couplings. Assume further that $S_1$ is fixed, and $S_2$ is free to move with respect to it along the x-axis. Under certain general assumptions on the dependence of the coefficients of coupling on position, it can be shown that for each value of the frequency f there exists a specific separation distance d = d(f) between $S_2$ and $S_1$, at which the mean mechanical force, evoked by the couplings between them, becomes zero. The system $S_2$ fluctuates around the equilibrium position defined by that separation distance. When the frequency of the current source f is changed, the mean ponderomotive force between the circuits changes, and $S_2$ moves to the corresponding, new region of fluctuational equilibrium. In general, the equilibrium position is a piecewise continuous function of the frequency, undergoing discontinuous jumps at certain critical values of f.

It is important to emphasize that the observed, constantly fluctuating motion of $S_2$ is not an accidental feature, but is essential to the process of constant transformation between electromagnetic and mechanical forms of energy, which maintains the coupled system in a stable configuration. This fact is underlined by another experiment carried out by the Doubochinski group [21]: If we prevent $S_2$ from moving along the x-axis, and instead allow it to pivot around an axis perpendicular to the x-direction, while the distance to $S_1$ remains constant, then $S_2$ goes into a rotary motion around its axis. That rotary motion fulfills the role of the back-and-forth fluctuational motion described in the preceding paragraph. In general, the coupled system adapts to external constraints by varying the frequency, phase and amplitude of its fluctuational motion.

Similar types of behavior can be demonstrated in experiments where both $S_1$ and $S_2$ are passive resonators, coupled with each other by inductive, resistive and capacitative couplings, while at the same time interacting with a third, active oscillating system S (for example a solenoid with a periodic voltage source, or a field of electromagnetic radiation) (Fig. 8).

It is a general property of multiply-coupled electromagnetic oscillatory systems, that the state of mechanical equilibrium is fulfilled only on the average. The constant fluctuations of the participating systems around certain average positions, in the form of either vibratory or rotational motions, play an essential role in the mechanism by which the coupled system maintains its stability.

Under certain conditions, it is possible to excite undamped stable oscillations of the resonators around their equilibrium positions, in which the frequency of the spatial (mechanical) oscillation can differ by one or two orders of magnitude from that of the electromagnetic oscillations in the circuits. In this case we obtain behavior analogous to that of the argumental pendulum. However, there is an essential difference between the argumental pendulum and the more complex phenomena connected with the grouping of multiply-coupled oscillators. In the case of the argumental pendulum, the mechanism of interaction could be interpreted in terms of analogies with phase-frequency-amplitude modulation in radio technology. Such analogies are no longer adequate in the case of multiply-coupled electromechanical systems, where it is necessary at the same time to take into account the interplay of pondermotive and electromotive effects.

Owing to technical and other limitations, the original experimental investigations of the spontaneous motion of multiply-coupled resonators, carried out in the 1970s, could only barely scratch the surface of what is surely a very rich and important domain of self-organizing phenomena. It would be of great interest to follow up on those investigations today, utilizing the technological possibilities have become available in the meantime.

**5. The participatory-generative character of argumental coupling**

The experimental and theoretical results summarized above, shed light on two extraordinary properties of argumental coupling of oscillators, which are absent from the conventional forms of coupling of oscillating systems:

(1) the existence of a discrete ("quantized") array of stable regimes of the coupled system;
(2) in each stable regime, the participating oscillators continue to function with very nearly their own original mean parameters, thereby retaining their "identities" within the coupled system. One could say that in the classical forms of coupling the component systems are *enslaved* to the collective regime, while in argumental coupling they freely *participate* in it, conserving (in the mean) their own individual parameters and peculiarities.

The first property we shall call "generative", the second "participatory". In this section shall examine both properties more closely, and show how the combined participatory-generative character of argumental interactions leads to a possible mechanism for generating an entire hierarchy of distinct physical objects, starting from a given set of oscillators.

For the purposes of illustration, we examine the argumental pendulum as a system created by the coupling two oscillating systems: (A) the pendulum, oscillating at or near its proper frequency under gravity and (B) the solenoid together with its high-frequency current source. In Section 2 we pointed out the qualitative difference between the pendulum's behavior for small amplitudes -- when it remains inside the interaction zone --, and its behavior at larger amplitudes, at which the argumental mechanism comes into play. This difference in behavior reflects a *fundamental difference in the type of physical coupling* between the two oscillating systems from which the total system is composed.

In the first case we have a classical situation of "forced oscillations under a periodic external force" in which the two systems are very far from resonance. Due to the large difference between the frequency of the alternating current and the proper frequency of the pendulum, very little energy is transferred from the field into low-frequency oscillations of the pendulum. Instead, the pendulum's natural oscillations are dampled out, and it becomes a "slave" of the electromagnetic field, executing forced oscillations at the frequency of the field. The original oscillatory regime (A) of the pendulum has ceased to exist, and we are left with a single, rigidly-coupled compound system (C). We could express the result of this classical form of coupling symbolically as follows:

A + B = C.

By contrast, in each of the stable regimes of the argumental pendulum the pendulum arm executes very nearly its natural periodic motion, while undergoing fluctuations in phase, frequency and amplitude as a result of its interaction with the field of the electromagnet. The generation of additional frequency components, provided for by the argumental mechanism, makes it possible for the low-frequency pendulum to exchange energy efficiently with the high-frequency field. The system of the electromagnet and its current source retains its basic frequency and amplitude characteristics, while at the same time experiencing periodic current fluctuations due to currents induced in the solenoid by the motion of the pendulum's permanent magnet. The oscillatory regimes A and B thus continue to exist in the compound system, so that the result of the coupling could be expressed symbolically as

A + B = { A, B, (AB)$_n$ }

where (AB)$_n$ represents one of the discrete series of stable regimes of the argumental pendulum.

We think the participation of the oscillatory systems A and B in the stable regime of the coupled system AB, is much more typical of the way real objects exist together in Nature, than the rigid forms of coupling characteristic of both classical and quantum physics. Nature works in a participatory manner. The very fact that the Universe is heavily populated by individual physical objects of all kinds, attests to principles of organization whereby such objects retain their individuality and integrity while at the same time interacting with each other to create larger objects, in which they then participate. We shall return to this observation in the following section.

We now turn to the *generative* potential of argumental interactions. Here the essential point is to recognize that each of the stable oscillation regimes of a system arising from the argumentally coupling of oscillating systems, deserves to be regarded as a *distinct physical individual* in its own right.

Consider, for example, an argumental pendulum oscillating in one of its stable quasi-stationary regimes. Such a regime is characterized not only by a definite mean amplitude and period, but also by a *cycle* of *exchange* and *transformation of energy* between the pendulum arm and the alternating magnetic field. In each half-period of the pendulum a certain definite portion of energy, corresponding on average to the frictional losses of the pendulum, is transferred from the magnetic field to the pendulum's motion. That portion ("quantum") of energy has been converted, in effect, from the frequency of the field to the much lower frequency of the pendulum, via the mechanism of phase-frequency-amplitude modulation described in Section 2 above. This cycle of transformation of energy is maintained over many periods (or indefinitely) and possesses an *active, self-regulating character*. One can observe, in fact, that the regime *actively* defends its stability and *actively* adapts and reacts to external influences, utilizing for this purpose a certain portion of its own energy flows (its "metabolic energy").

Clearly, a *functional regime* of this sort signifies something different from a *material object* in the everyday sense of the word. We consider, however, that a "something" which has definite physical parameters, that exists on the basis of a self-regulated flow of energy, that actively maintains itself, adapts and reacts to external conditions, deserves to be regarded as a *real physical object*. Naturally this reasoning applies not only to the argumental pendulum, but to the stable regimes of argumentally-coupled oscillators in general.

From this standpoint argumental coupling shows itself to be a powerful instrument for the generation of individual physical objects. Suppose we have a set of oscillating systems A, B, C, D, E, ... which are capable of entering into argumental interactions with each other. The argumental coupling of any pair of them produces an array of stable regimes, each of which constitutes an individual physical object. Thus, from A and B we can obtain objects $(AB)_1$, $(AB)_2$,...,$(AB)_n$,... etc., from C and D the objects $(CD)_1$, $(CD)_2$,...,$(CD)_m$, ... and so forth. Each of those objects, as a stable oscillatory regime with specific frequency, amplitude and phase characteristics, represents an oscillating system which is in principle capable of entering into argumental couplings with other oscillating systems. For example, the argumental coupling of $(AB)_n$ with $(CD)_m$ generates an array of individual physical objects $[(AB)_n(CD)_m]_1$, $[(AB)_n(CD)_m]_2$, .... $[(AB)_n(CD)_m]_k$ ... This process can in principle be applied again and again, leading potentially to gigantic numbers of individual physical objects, related to each other in a hierarchical manner. The new objects obtained at each stage form the basis for generating the next level of objects. Due to the participatory nature of argumental coupling, each newly-formed object maintains its individual existence and characteristic parameters, while participating in the formation and life of objects on higher levels of the hierarchy.

Naturally, the realization of such hierarchies of oscillatory regimes in concrete physical systems, can be limited by a variety of factors and conditions. Of particular interest are the conditions of stability of the functional regimes at different levels of the hierarchy, and the effects of "quantum jumps" in the regimes of participating systems.

Without entering into details of a concrete physical system, consider for example the effect of argumentally coupling two stable regimes, say $(AB)_3$ and $(CD)_5$, to obtain new objects of the type $[(AB)_3(CD)_5]_k$. The existence of any stable regime of the coupled system $(AB)_3(CD)_5$ clearly presupposes that the corresponding phase-frequency-amplitude fluctuations of the combined system must remain within the zones of stability of each of the participating systems $(AB)_3$ and $(CD)_5$. In case internal or externally-imposed fluctuations cause a system to "jump" to a different stable regime (for example $(AB)_3 \rightarrow (AB)_1$), this can trigger a chain of abrupt transitions in functional regimes of all the systems in which $(AB)_3$ participated, propagating upward in the hierarchy. Cascades of an analogous sort play an essential role in the mechanism of control of many natural processes, including especially in living organisms.

These brief remarks are intended only to give a preliminary glimpse of a vast domain of new oscillatory phenomena, which is opened up by the "participatory-generative" properties of argumental interactions. This domain has only barely begun to be explored experimentally, with the observation of "higher-order" regimes involving two or more pendula maintained in stable regimes by argumental interaction with a single electromagnet. Here is a rich field for future research.

**6. Physical objects and their interactions**

Having reviewed results from the experimental and theoretical investigations of argumentally-coupled systems, we are now in a position to present some of our ideas concerning the foundations of physics. We shall not propose any grand formal theory, but rather *a new way of looking at physical objects and their interactions*, differing in important respects both from the viewpoint of classical physics and from that of modern quantum theory.

The study of Nature presents us with a seemingly inexhaustible number and variety of *individual physical objects*: galaxies, stars, planets, living organisms, molecules, nucleii, elementary particles etc., existing on widely-differing scales of space and time, and manifesting a character of wholeness and invariant features that lead us to regard them as distinct individuals. We observe a marked tendency on all scales, for physical individuals to interact with each other, to associate together in more or less stable groupings and to participate in the formation of larger objects having their own individual characteristics.

Science has not yet developed a unified approach to the origin, stability, interaction and "social behavior" of the many species of physical objects populating our Universe. Instead, one encounters widely divergent ideas and explanations, depending on the scientific discipline and the type of object involved.

One might argue that the reason for the great differences in conceptual approaches lies in the nature of the objects themselves, and the fact that they pertain to *different levels of organization* of physical reality. The authors, however, find it difficult to believe that the Universe would operate in a fundamentally different way on one level of organization, than on another. It seems far more likely that *the underlying principles of generation, maintenance and "social" interaction of physical objects in the Universe are everywhere very much the same;* and that the common principles have remained hidden owing to the lack of development of a suitable unified conceptual framework, and to certain habits of thinking passed down from classical (Newtonian) physics.

The discovery of argumental interactions and their generative-participative properties suggested to the authors a possible pathway towards the solution of this problem.

The first step is to adopt a general dynamical notion of what should be meant by the term "individual physical object" -- a notion along the lines we suggested in our discussion of the stable regimes of argumentally-coupled oscillators in the preceding section, and applicable to all levels of organization of the Universe. The decisive criterion, is that an individual physical object must be conceived always as something inseparable from a specific "regime of functioning", i.e. from a specific, active physical process by which the object maintains itself in a stable manner, interacting with and reacting to changes in its environment while retaining its essential characteristics. To put it more directly: *the real individual objects are for us the functional regimes themselves.* To the extent a system such as a star -- or a living cell -- can exist in various different stable or quasi-stable regimes, each of those regimes constitutes for us a *distinct object, a distinct physical individual*. By their very nature, functional regimes invariably involve cycles of flow and transformation of energy, and are thus oscillatory in nature.

The stable modes of the argumental pendulum provide the most transparent examples and models of distinct physical individuals in our sense. Here the details of the functional regimes and the mechanism by which they are "born", are most easily accessible to study.

In the case of the objects of Nature, we often have only limited knowledge of the functional regimes and their interrelations, and cannot always precisely distinguish different regimes clearly from one another. Nevertheless, it is a matter of very general observation that natural systems are found in distinct stable or quasi-stable dynamic states; that these states -- as individual physical objects in our sense -- can be recognized and classified in discrete categories according to their characteristic features; and that the transitions between such states, where they are possible, tend to be more or less abrupt and jump-like. In each case in which we are able to discern the internal structure of a physical object, we find that it is constituted from the interaction and coupling of other physical objects, which participate in it while maintaining their own individuality. Conversely, individual physical objects of every known species, up to at least the level of galaxies, are found to participate in some sort of larger objects (e.g. galactic clusters, superclusters). The transition of a physical system from one stable regime to another, can trigger a cascade of changes in the entire hierarchy of physical objects in which the given stable regime participates.

In all these respects we find a broad resemblance between the organization of physical objects in the Universe, and the hierarchies of stable oscillatory regimes generated by argumental couplings. This analogy becomes still closer when we adopt a criterion for what should be meant by the term *physical interaction*, which is consistent with the dynamic concept of "individual physical object" adopted above. A *true physical interaction* must be conceived of as a real dialog between individual physical objects: a process involving constant exchanges of energy, in which each object accommodates its functional regime to that of the other, without either of them losing its essential identity. When such a process of interaction evolves into a self-regulating, quasi-stationary regime, we speak of a coupling of the objects and of the birth of a *new physical object* as a result of that coupling.

Argumental interactions, as embodied in the argumental pendulum and in the electromotive/ ponderomotive interaction of electrical oscillators described in section 4, provide the most transparent models for this notion of interaction and coupling.

Considering these matters from a fundamental point of view, we can hardly doubt that the emergence of individual physical objects, on the one hand, and of interactions among physical objects, on the other, represent two complementary aspects of a single physical reality. The very same functional regime, by which a physical object maintains its individuality and identity, is at the same time the basis for its interactions with other objects. The functional regime of any physical object involves the entire Universe directly or indirectly. Hence the functional regimes of all objects are constantly reacting to and accommodating to each other. This, we propose, is the ultimate source of the effects classical and quantum physics attribute to "fundamental forces" acting between the particles of matter.

The authors are perfectly aware that their proposed way of looking at physical objects and physical interactions may appear rather paradoxical, at first sight, and raises many questions. In closing we shall address only one of the most important of these, permitting ourselves at the same time to add some historical and methodological comments which may help clarify what has been said above.

Among the most important questions, is whether current empirical knowledge concerning atoms, electrons and other microphysical entities, justifies their being considered "individual physical objects" in our proposed dynamic sense. This question is immediately connected with the problem of the ultimate origin of gravitation and the other "fundamental forces of physics." The difficulty here is perhaps less one of empirical evidence, than it is habits of thinking going back to classical (Newtonian) physics, which regards "the elementary constituents of matter" as entities existing somehow in and of themselves, without requiring any activity or regulatory functions for their maintenance and stability.

This essentially static notion of the constituents of matter goes hand in hand with another drawback of classical physics: the lack of any intelligible explanation for the existence of *interactions* between physical objects. Indeed, without acknowledging any *activity* intrinsic to the existence and maintenance of physical objects, it is hardly possible to understand how physical objects could exert forces on each other. Classical physics avoids this problem by simply *postulating* the existence of "fundamental forces" such as gravitation, having no intelligible basis in the nature of the objects upon which they act. Although Newton himself expressed his dissatisfaction with this practice, it has had a deep influence on the habits of thinking of physicists up to the present day.

On the other hand if – as Newton's contemporary Leibniz had argued -- the essence of any physical object lies in a constant activity, extending implicitly to the entire Universe, then the existence of interactions between objects is no longer mysterious. Forces would then be a secondary effect of the dialog between the functional regimes of the physical objects, in the way we have suggested. But in order to realize this idea in the form of a truly dynamic theory of physical interactions, it would be necessary to know much more about the functional regimes of physical objects, including especially the microphysical particles that constitute material bodies. What does present-day science have to say, for example, concerning the functional regime of an electron? At first glance nothing at all. But the answer depends on how one interprets the evidence of quantum physics.

At the outset of the development of quantum physics Louis de Broglie recognized that the existence of an electron must somehow be inseparably connected with a *high-frequency* *oscillatory process, extended in space*. This and subsequent successes of Schrödinger's wave mechanics, should have been seen as a vindication of Leibniz's dynamical standpoint and a first step toward clarifying the functional regimes underlying microphysical objects Unfortunately, the early attempts by Schrödinger to develop an electromagnetic interpretation of the wave function were abandoned, as was also the interesting attempt by Nernst to understand quantum phenomena (and gravitation!) as a product of interactions with an oscillating medium [3]. The elaboration of quantum mechanics took a completely different direction. Lacking an intelligible notion of how quantization occurs *both* in microphysical systems and in macrophysical systems such as the solar system, the introduction of the quantum of action as a postulate of quantum mechanics led to a schism in the physical picture of the world. The prospect for understanding electrons, protons and other microphysical entities as essentially *dynamic* objects -- and thereby also of grasping the origin of the fundamental interactions between them -- receded into the future.

We think the time may now be ripe to reconsider these matters. We believe the example of argumental interactions and the ideas sketched above, while not pretending to provide concrete answers now, may nevertheless prove to be a fruitful starting-point for a more coherent understanding of physical objects and their interactions, independent of the scale and level of organization in the Universe.

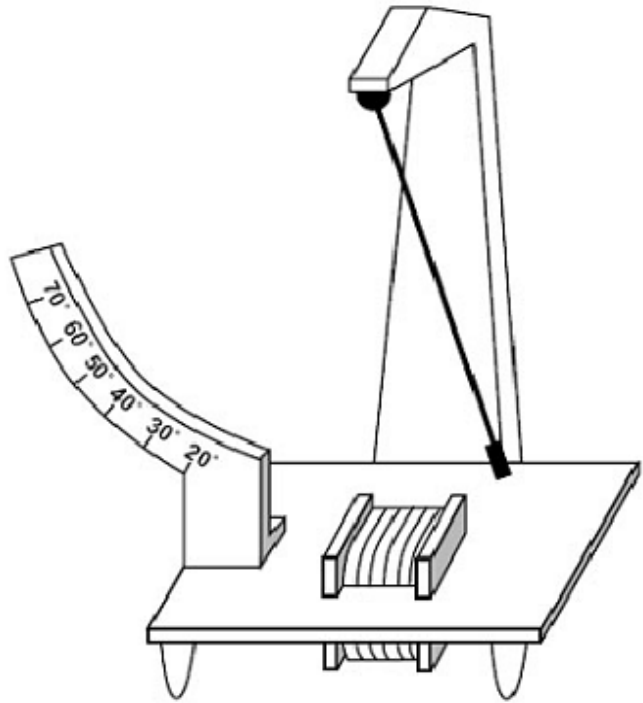

Figure 1 – The argumental pendulum

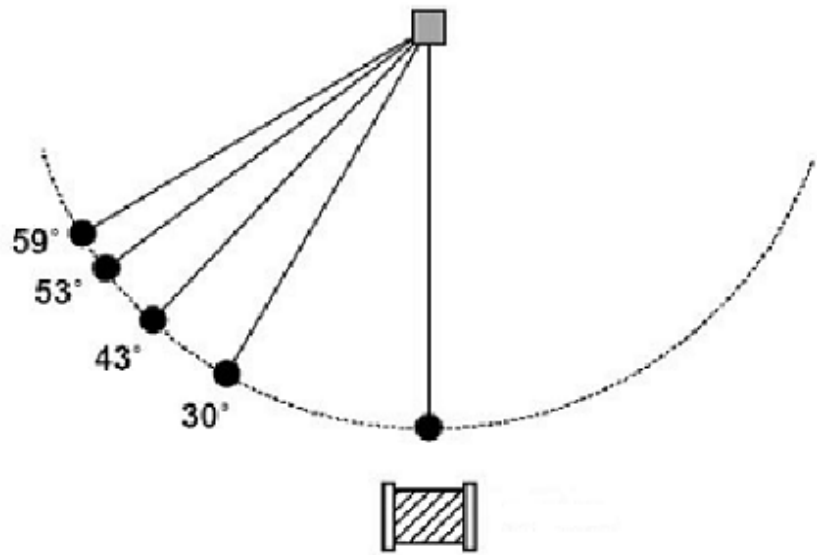

Figure 2 – Discrete amplitudes of the argumental pendulum

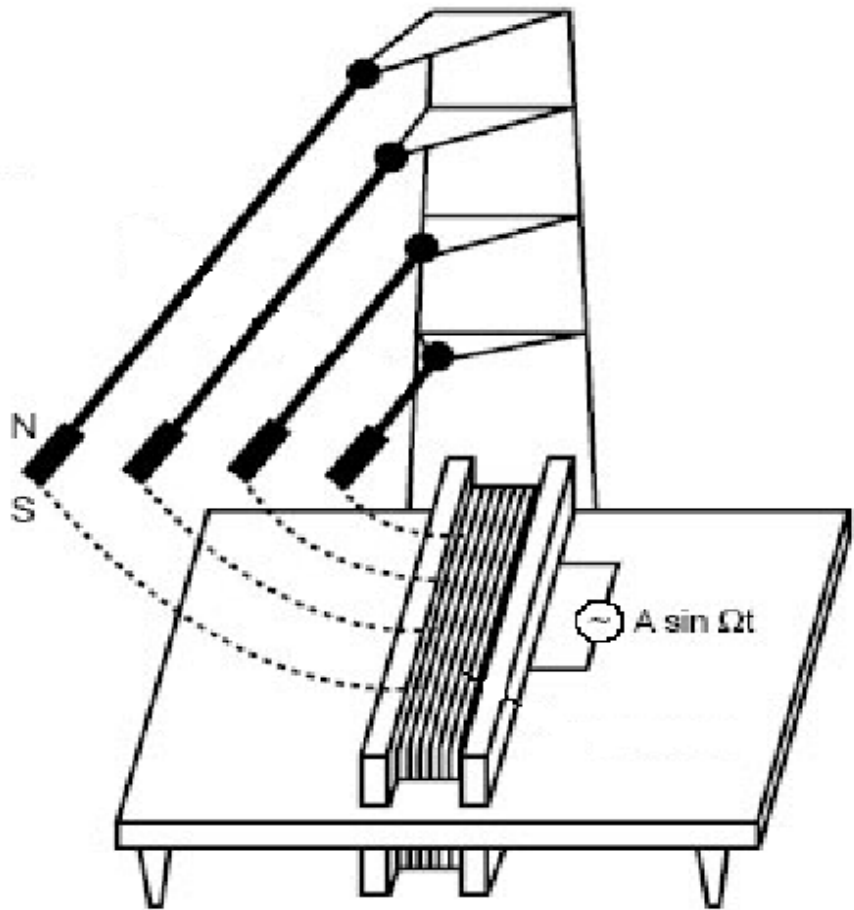

Figure 3 – Multiple argumental pendula "fed" by interaction with a single high-frequency source. The pendula can operate at different frequencies, each at any one of its own discrete series of "quantized" amplitudes.

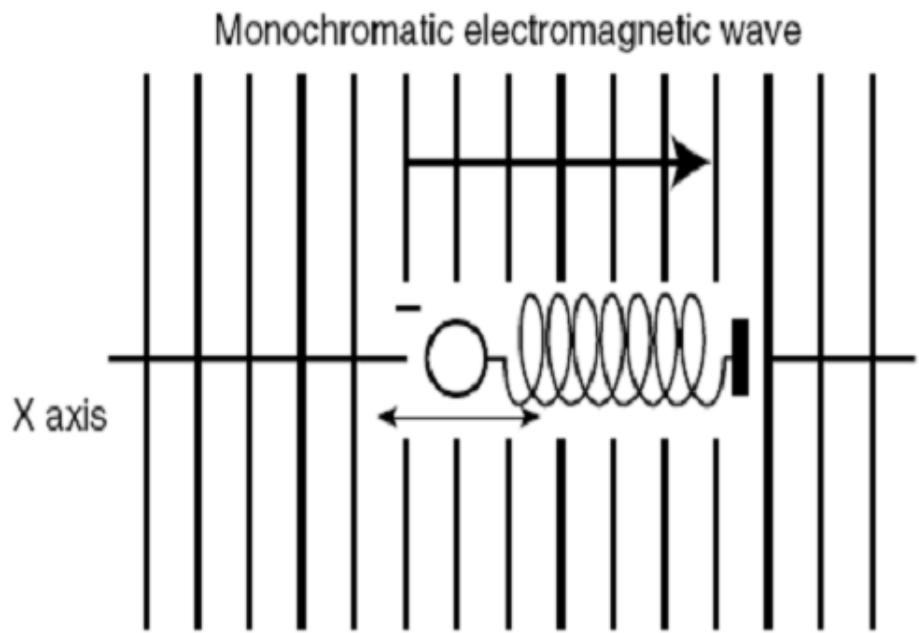

Figure 4 – The argumental analog of Planck's "elementary oscillator"

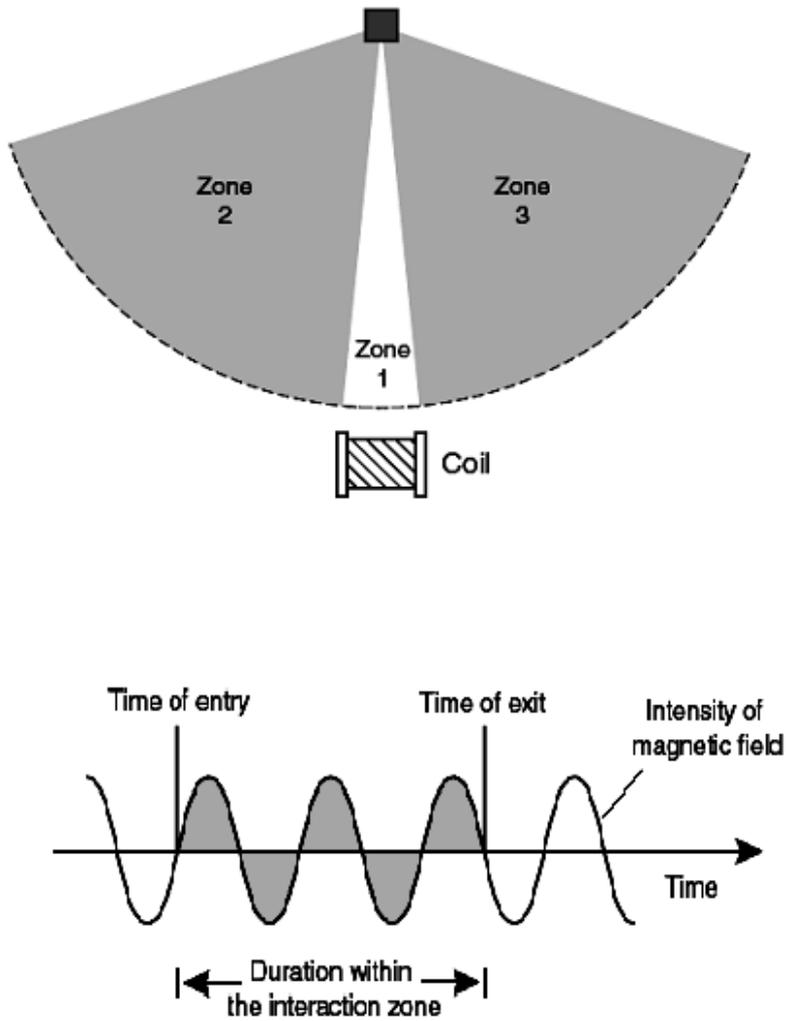

Figure 5 – The effect of a single passage of the pendulum through the "interaction zone" (Zone 1 in the diagram) depends on the relationship of the phases of the magnetic field at the moments of entry and departure from the zone. In the example shown here, the pendulum has left the zone after a non-integral number of cycles of the field, thereby experiencing a net acceleration.

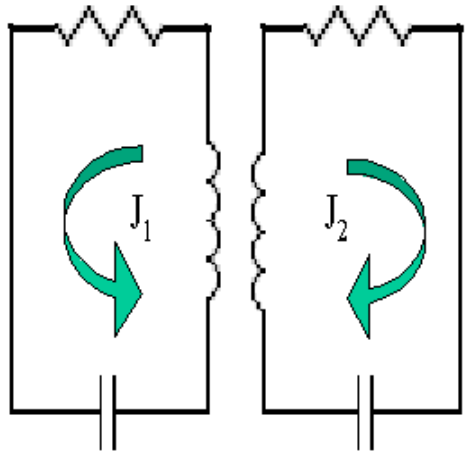

Figure 6 – Inductively coupled circuits exert a force on each other proportional to the product $J_1 \times J_2$

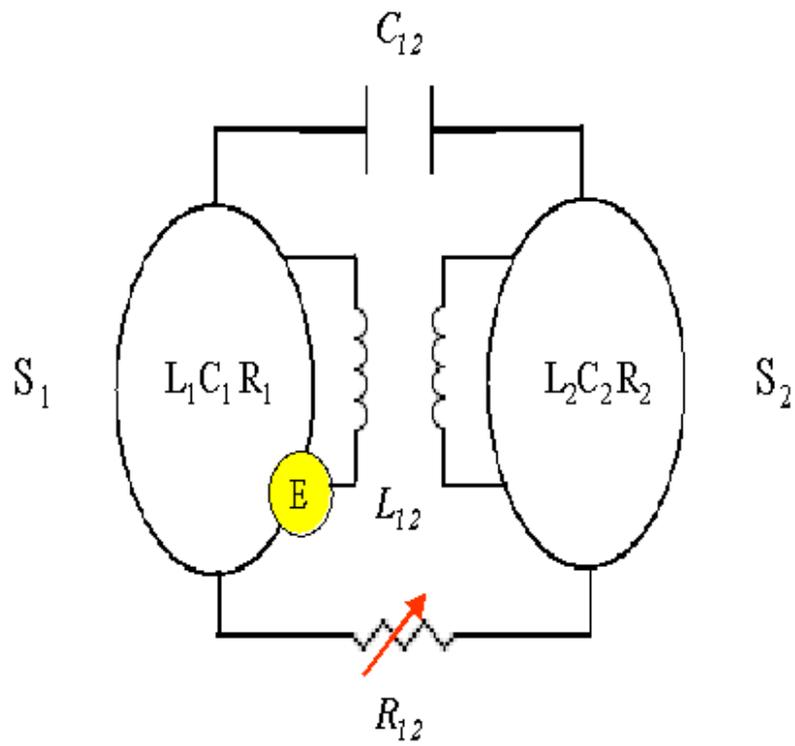

Figure 7 – Schematic of experiments with multiply-coupled oscillating systems, free to move along the horizontal axis.

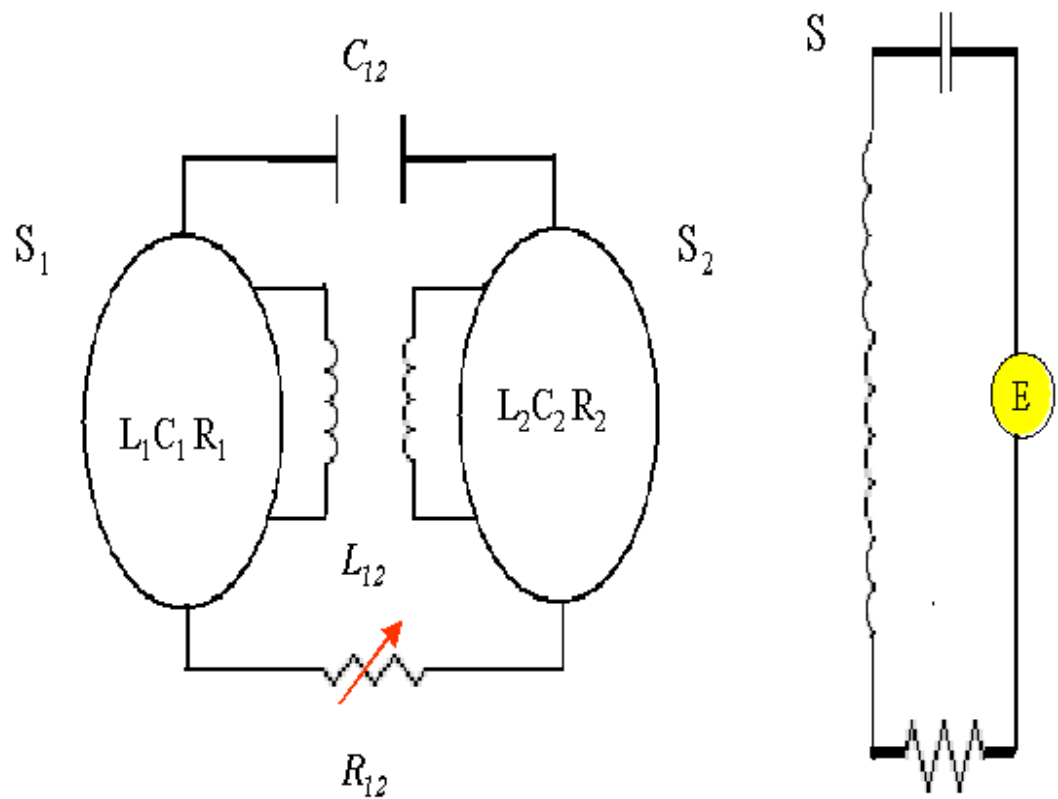

Figure 8 – Schematic of experiments with multiply-coupled resonators